\documentclass{llncs}
\usepackage[margin = 1.15in]{geometry}
\usepackage[utf8]{inputenc}
\usepackage{amsmath}
\usepackage{amssymb, amsfonts, cite, authblk}
\usepackage{ifthen}
\usepackage[boxed, linesnumbered]{algorithm2e}
\usepackage{comment}
\usepackage{url}
\usepackage{float}

\usepackage{pgf}
\usepackage{tikz}
\usetikzlibrary{arrows,automata}

\usepackage{microtype}

\usepackage{hyperref}
\usepackage{multirow, multicol}
\usepackage{subfigure, graphicx} 
\usetikzlibrary{arrows,shapes}

\tikzset{circle/.pic={
\node[circle, aspect=1, draw, minimum size=0.3cm, text width=0.2cm] () at (0,0) {\tikzpictext};
}} 
\usepackage{natbib}
\usepackage{microtype}
\usepackage{pgfplots}
\usepackage{booktabs}
\pgfplotsset{compat=1.10}

\bibpunct{[}{]}{;}{a}{,}{;}
\begin{document}

\title{Routing Games in the Wild: Efficiency, Equilibration and Regret}
\subtitle{Large-Scale Field Experiments in Singapore}
\date{}

\author{Barnab\'e Monnot\inst{1} \and Francisco Benita\inst{2} \and Georgios Piliouras\inst{3}}
\institute{Engineering Systems \& Design, Singapore University of Technology and Design \email{monnot\_barnabe@mymail.sutd.edu.sg} \and Engineering Systems \& Design, Singapore University of Technology and Design \email{francisco\_benita@sutd.edu.sg} \and Engineering Systems \& Design, Singapore University of Technology and Design \email{georgios@sutd.edu.sg}}

\maketitle

\begin{abstract}
Routing games are amongst the most well studied domains of game theory.
How relevant are these pen-and-paper calculations to understanding the reality of everyday traffic routing?
We focus on a semantically rich dataset 
that captures detailed information about the daily behavior of thousands of Singaporean commuters
and examine the following basic questions:

\begin{itemize}
  \item Does the traffic equilibrate?
  \item Is the system behavior consistent with latency minimizing agents?
  \item Is the resulting system efficient?
\end{itemize}

In order to capture the efficiency of the traffic network in a way that agrees with our
everyday intuition we introduce a new metric, the \textit{stress of catastrophe}, which reflects the combined inefficiencies of both tragedy of the commons  as well as price of anarchy effects.







\end{abstract}

\section{Introduction}
Congestion games are amongst the most historic, influential and well-studied classes of games.
As the name suggests, they were designed to capture settings where the payoff of each agent depends on the resources he chooses and how congested each of them is.
Proposed in \citep{rosenthal73} and isomorphic to potential games \citep{potgames} (in which learning dynamics equilibrate), they have been successfully employed in a myriad of modeling problems. Naturally, one application stands above the rest: modeling traffic. Having strategy sets correspond to the possible paths between source and sink nodes in a network is such a mild and intuitive restriction that routing/congestion games are effectively synonymous to each other and jointly mark a key contribution of the field of game theory.

Routing games have also played a seminal role in the emergence of algorithmic game theory.
The central notion of Price of Anarchy (PoA), capturing the inefficiency of worst case equilibria, was famously first introduced and analyzed in routing games \citep{KoutsoupiasP99WorstCE,roughgarden2002bad}. Routing games have set the stage for major developments in the area such as the introduction of regret-minimizing agents 
\citep{pota}  that eventually led to the consolidation of most known PoA results under the umbrella
of  $(\lambda,\mu)$-smoothness arguments \citep{Roughgarden09}. Impressively, this work
  established that PoA guarantees are robust for a wide variety of solution concepts such as correlated equilibria,
regret-minimizing agents and approximations thereof.
Finally, congestion games still drive innovation in the area with results which extend the strength and  applicability of PoA bounds for large routing games  \citep{feldman2016price}, as well as dynamic populations
\citep{lykouris2016learning}.

With every successive analytical achievement seemingly chopping slowly away the distance between theoretical models and everyday reality, the PoA constants for routing games, e.g., the $4/3$ for the nonatomic linear case \citep{roughgarden2002bad} have become something akin to the universal constants of the field.
Small, concise, dimensionless, they seem almost by their very nature to project purity and truth.
But do they? After all, there are many of them. In the case of quadratic cost functions PoA
$\approx1.626$,
whereas for quartic functions, which have been proposed as a reasonable model of road traffic, PoA
$\approx 2.151$ \citep{Sheffi1985urban,RoughgardenBook16}.
What do these ``small constants'' mean in practice? Quite a lot. An increase of inefficiency from $4/3$ to
$2.151$ in Singapore would translate in the loss of approximately 730,000 work hours \textit{every single day}.
Do any of these ``back-of-the-envelope'' theoretical calculations have any predictive power \textit{in practice}?

At the antipodes of the aforementioned theoretical work, other, similarly recent
theoretical approaches hint that PoA analysis might actually not be reflective of the realized behavior in real networks. One type of work focuses on the instability of worst case equilibria, e.g.,  \citep{Kleinberg09multiplicativeupdates,ITCS15MPP}. Specifically, \citep{panageas2016average} show that although bad equilibria may exist, an average case  analysis which ``weighs'' each equilibrium proportionally to its region of attraction typically reveals a picture that is much closer to optimal than PoA analysis. So, PoA analysis may be over-pessimistic.
Distressingly, \citep{colini2016price,scarsini2017poa} argue something orthogonal, which at first glance appears rather counterintuitive. They argue that networks with low PoA, e.g. PoA\(=1\), which are typically considered optimal, might actually reflect traffic flows which are deadlocked in severe traffic jams. \footnote{Indeed, if we keep increasing the total flow in e.g. Pigou's example, eventually both in the optimal and equilibrium flows almost all flow will be routed through the slow link. In layman's terms, if you keep increasing the number of cars in a city eventually the traffic will be so bad that you might as well walk to your destination.}
Finally, PoA calculations can be invalidated if we move into theoretical models that allow for risk averse agents \citep{Piliouras:2016:RSP:3007189.2930956, NikolovaS14-OR, Angelidakis:2013, HotaGS13, Meir:2015:PWG:2847220.2847242,nikolova2015burden}.
So, summing up, we have that a high PoA might be artificially large, whereas a low PoA might actually reflect socially undesirable (in fact catastrophic) conditions. Can we then trust these numbers to begin with?

At this point, as theory alone does not suffice to provide a definitive answer, it makes sense to examine some real world networks at a fine level of detail.




\textit{Our goal} is to perform the first-to-our-knowledge game theoretic modeling and investigation of a real world traffic network (specifically Singapore's traffic network) based on repeated large scale field experiments with thousands of participants (specifically focusing on Singaporean students\footnote{For a discussion on the representativeness of our sample see sections \ref{sec:datadesc} and the Appendix.}).  Our dataset includes detailed information that allows us to inspect minute-by-minute\footnote{In fact at a finer precision than that. See section \ref{sec:datadesc} for a detailed description.} the concurrent decision making of thousands of commuters, as they respond and adapt to traffic conditions.

We focus on arguably the three most basic questions: First, is the system at equilibrium? Second, is this equilibrium consistent with the classic hypothesis of latency minimizing agents? Finally is the resulting system efficient? Before we explore the answers to these questions as provided by the data, let's try to disambiguate the questions themselves.

\textit{Is the system at ``equilibrium''?} Here, we should clearly point out that by equilibrium we mean the formal mathematical notion of equilibrium, i.e. a stationary point. At the first level of inspection, we are not concerned with whether the outcome that the system equilibrates upon is necessarily stable in a game-theoretic sense. We are merely asking ``are the agents continuously adapting their behavior from day-to-day'' (i.e. the paths they choose, the modes of transportation, and so on)? If significant number of agents choose the same actions from day-to-day this would indicate that the system has indeed reached a fixed point (stasis) and furthermore that at this stable system state there is little entropy/randomness.
Such a result is consistent with best response and best response dynamics, with the instability results of mixed Nash equilibria for multiplicative weight update algorithms \citep{Kleinberg09multiplicativeupdates,ITCS15MPP,panageas2016average} as well as
with some other concurrent dynamics (e.g. imitation dynamics) \citep{Ackermann:2016:CID:2931748.2931779,Fotakis08}. However, it is not a universal consequence of no-regret learning in congestion games, which can be persistently random or adaptive, leading to spiky congestion patterns with high makespan \citep{pota}.

\textit{Is the equilibrium ``economically stable''?}  Naturally, from a game theoretic perspective, we wish to understand whether the resulting equilibrium is a Nash equilibrium (or at least if in the case of adapting agents most have low regret when comparing their performance with the best path in hindsight). For a real traffic network, however, it is not practically feasible to compute true ``best responses'', since there is an astronomically large number of paths to consider as possible deviations and we do not have data on all paths. Moreover, truly optimal paths might consist of complicated combinations (metro for the first half of the route, then a specific car route for the next half) that no single agent has figured out.
Instead of considering the typical regret of each agent, we instead  estimate inefficiencies at the individual level by quantifying the daily empirical ``imitation'' regret for each agent, i.e., how much faster could each agent have reached their destination if they had clairvoyant access to all the traffic information from our dataset and chose the best possible route with hindsight\footnote{Using variants of the standard notion of regret, which are easier to compute in practice has been the subject of recent theoretical work especially in learning in auctions. See ``envy'' defined in \cite{daskalakis2016learning}}. Effectively, we are comparing each student to the best one with hindsight in their cluster/neighborhood (going to the same school) and hence the term imitation-regret.

\textit{Is the system  ``efficient''?}  Traditionally, ever since its inception, the notion of Price of Anarchy has been considered the gold standard for system efficiency with a low PoA considered equivalent to system optimality. The results in \citep{colini2016price,scarsini2017poa} in which hopelessly deadlocked traffic jams score perfect PoA scores point out a clear dichotomy between what PoA analysis identifies as an efficient traffic network and what we in our everyday experience identify as a well-functioning network.
The reason for this divide lies on the fact that PoA analysis completely disregards any  inefficiency that is connected to tragedy of the commons effects. The degree of exploitation of common resources\footnote{Or how many people decide to buy cars and hence consume significant resources from the public good (i.e. the road network)?} is seen as an exogenous parameter that is not subject to strategic considerations and which is independent to the inefficiencies introduced by PoA effects. However, without studying the combined inefficiencies from tragedy of the commons as well as PoA we become blind to the possibility of systems  turning into optimal-PoA catastrophes. Distressingly, as PoA decreases with the increase of the total load (converging to $1$) \citep{colini2016price,scarsini2017poa} it encourages increasing participation, thus forcing the system to keep marching towards the high congestion regime and effectively a system catastrophe.

We wish to shine a light on this unexplored but critical phenomenon and for this reason we define a new metric of system inefficiency that is defined as the ratio of the social welfare at equilibrium divided by the optimal social welfare (when we discount for congestion effects). That is although the numerator is as in the PoA (in the case of non-atomic flows), the denominator is computing the average social ``blue-sky'' optimal welfare as follows: Each agent imagines the scenario where she alone was in the network and computes the best of minimum length/latency for herself. We argue that this makes sense from an everyday experience perspective, as the average user naturally incurs the latency of the path she uses (numerator summand) but also has an intuitive grasp of how long it would take to cover this distance if the externality costs imposed by the other users where removed. We call this ratio, the \textit{Stress of Catastrophe (SoC)}. As this ratio grows (regardless of the value of PoA) the system's long term persistence and health is jeopardized. Practically successful networks should have small SoC, which implies small PoA but not the other way around.

\subsection*{Result Snippets}

\begin{itemize}
    \item We show that most students use the same means of transportation across trips and that a large number of them consistently selects the same route. For example, when controlling for students who use consistently the same means of transportation across different days, the percentage of subjects selecting the same route is very high, in the order of 94\%.  (see Section \ref{sec:consistency}).
    \item The empirical regret distribution has a median value of 4 minutes 40 seconds and mean approaching 6 minutes for an average travel time of around 29 minutes (see Section \ref{sec:regret}).
    \item Finally, we define and estimate the Stress of Catastrophe at 1.34, with marked contrast when discriminating by mode of transportation (see Section \ref{sec:poa}). These findings are shown to be  consistent across different days.
\end{itemize}

\section{Description of the Data}
\label{sec:datadesc}

We focus on a semantically rich dataset from Singapore's National Science Experiment (NSE), a nationwide ongoing educational initiative led by researchers from the Singapore University of Technology and Design (SUTD)~\citep{Monnot2016,Wilhelm2016} (more details to be found in Appendix \ref{sec:app-nse}). This dataset includes precise information about the daily behavior of tens of thousands of Singapore students that carry custom-made sensors for up to 4 consecutive days, resulting in millions of measurements. Indeed, every 13 seconds, the sensor is able to accurately log its geographical location as well as other environmental factors such as relative temperature and humidity or noise levels.

The students are dispersed throughout the city-state and their daily commutes to school are reasonably long for them to meaningfully interact and experience the daily traffic. For this reason, we focus on the morning trip they undertake to reach their school from their home. To guide the reader unfamiliar with Singapore's road and public transport network, we give a brief optional introduction in Appendix \ref{sec:app-singapore}.

\begin{figure}[!ht]
	\centering
	\begin{minipage}[c]{0.46\linewidth}
        \centering
        \includegraphics[width=0.95\linewidth]{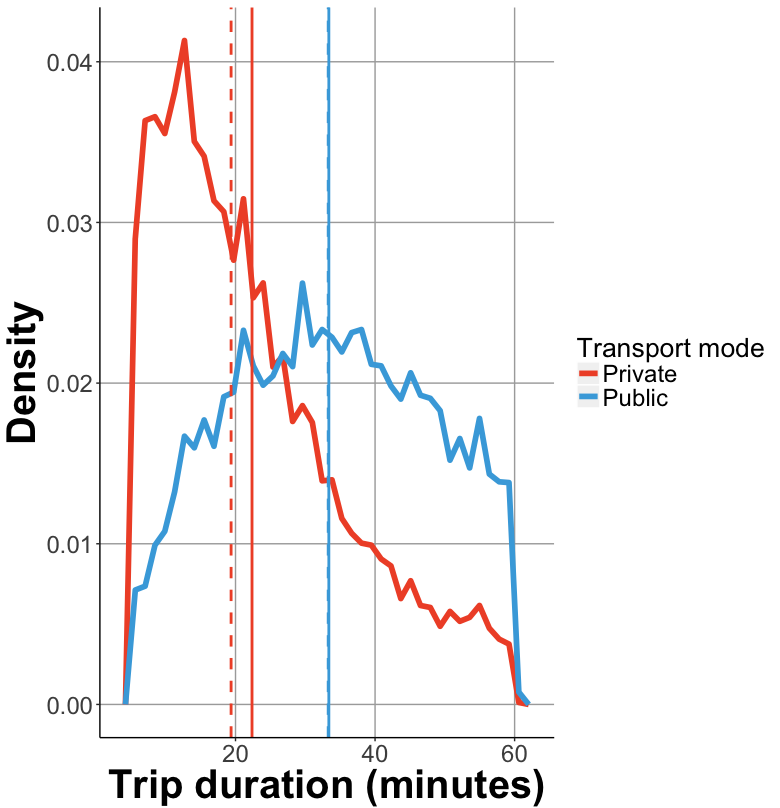}
	\end{minipage}
	\hspace{0.05\linewidth}
	\begin{minipage}[c]{0.46\linewidth}
		\centering
		\includegraphics[width=0.95\textwidth]{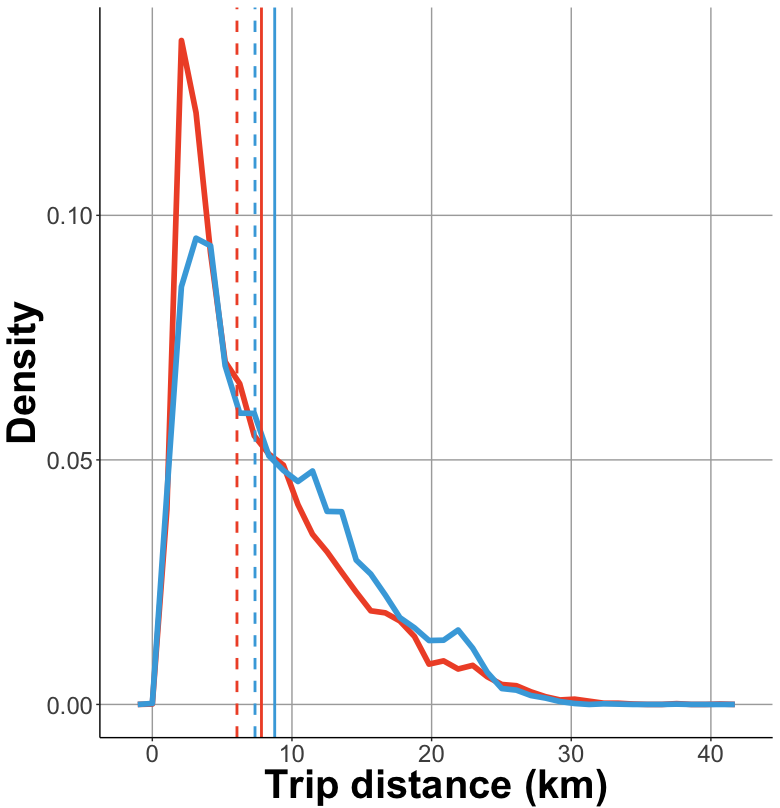}
	\end{minipage}
	\caption{\textit{Left:} Density plots of trip durations per mode. We note that car trip durations are typically short and more concentrated around a peak value of 15 to 20 minutes, while public transportation trip durations are scattered between 20 to 50 minutes. \textit{Right:} Density plots of trip distances per mode. Surprisingly, there does not appear to be any difference between the two lines, indicating that distance does not factor in the choice of transportation mode. Median is represented by a dashed line and mean by a solid line.}
	\label{fig:duration}
\end{figure}

The mode of transportation chosen by the students can be identified using accurate algorithms, e.g. car (driving or being driven to school) versus bus or metro, estimate source and sink destinations (focusing on home-school pairs) as well as their mode-dependent available routes. Some descriptive charts are given in Figure \ref{fig:duration} relating the durations and distances traveled for private and public transportation trips.

\subsection*{Representativeness of the Sample}

A word on the experiment's sample: students may be a restricted class of residents, but we argue that it however provides a tangible idea of Singapore's mobility. First, as of 2015, the size of the student population up to Pre-University level totals about 460,000 residents. In contrast, the active population's size, as of 2015, is above 2.2 million\footnote{Statistics were compiled from data.gov.sg}. Our clean dataset comprises 15,875 unique students, distributed between the three main type of institutions in Singapore (Primary, Secondary and Pre-University).

For the purpose of our study, it may also be noted that most of the analysis does not require a complete sample of the population. Students in private transportation experience the same level of congestion as their peers and active individuals, hence estimates over their population translate to estimates over the whole of Singapore's mobility users. It is even more true for students in public transportation: their trips are possibly the same as those of the active population. Indeed, we find that the ratio of public to private transportation users in our sample closely mirrors that of the population as a whole\footnote{Household Interview Travel Survey 2012: Public Transport Mode Share Rises To 63\%, LTA News Release}, as 57\% of students in our dataset use public transportation.

\begin{figure}
  \centering
  \includegraphics[width=0.6\linewidth]{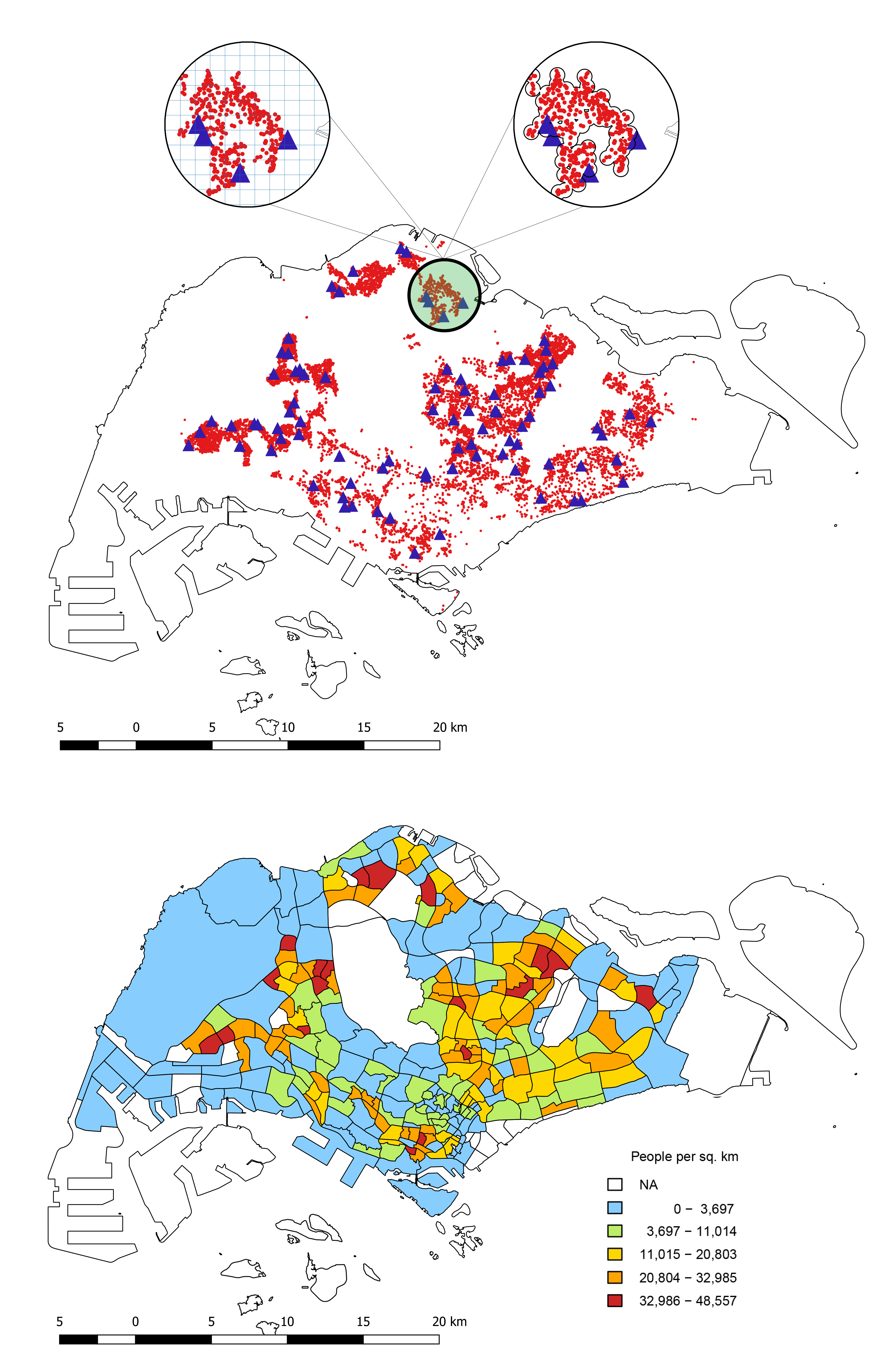}
  \caption{\textit{Up:} Home locations (red dots), school locations (blue triangles) and spatial clustering methods, discussed in Sections \ref{sec:regret} and \ref{sec:meth-imitregret}. \textit{Down:} Density map of Singapore. Blue areas are less populated while red areas are denser.}
  \label{fig:clusters}
\end{figure}

As shown in Figure \ref{fig:clusters}, the home location sample is geographically distributed, so as to not focus on a particular area of the city. On the other hand, the distribution of schools may not reflect endpoints of trips made by the active population. As an example, it can be observed that few schools are located in the city center, which houses a large number of office buildings. This constitutes one limitation of our dataset, perhaps softened by the fact that active population and students may still share parts of their trip together close to the residential area, and thus experience the same congestion.

As a measure of the hardness of running field experiments, it is useful to compare with the well-known sociological field experiments such as Miglram's small world experiment~\citep{travers1967small}. In this famously successful experiment, only 296 letters were sent, the vast majority of which (232) never reached their destination and out of the 64 letters that eventually did reach the target contact, only the average path length was around five and a half or six, which eventually generalized to universal statements about the connectivity of human networks. In comparison, we perform field work in a particularly noisy field -- transportation -- but in our case we have acquired several orders of magnitude more clean data. Previous studies in Singapore have also met limitations of different sorts, e.g the works of \citep{Sun2012, Lee2014, Holleczek2015} or \citep{Poonawala2016}, are either based on smart card data -- focusing only on public transport commuter trips -- or GSM (cellular phone) data.


\section{Findings}

\subsection{Equilibration and Empirical Consistency}
\label{sec:consistency}

If we believe that our system is at equilibrium, then we should expect that the students' route decisions do not vary wildly between successive days of study. We investigate the issue from three different angles. First, we compare the modes of transportation selected by one student over the days of the experiment. Second, we improve the previous result by considering whether the selected routes are identical (e.g. always use the same combination of bus and train, or always use the same road on car). Third, building on our geographical clustering method described in the following Section, we investigate the question of whether the fastest student in the cluster on one day remains the fastest over all days of experiment.

The first analysis shows that more than 60\% of students have used the same principal mode of transportation in all morning trips available in our dataset. We are here discriminating between trips where the principal mode of transportation is either the train, the bus or the car. We define as principal the mode with which the student has traveled the longest distance. The fraction increases to close to two thirds (65\%) of the samples if we simply discriminate between the students using public transit from those who use private transportation.

For the second analysis, we have implemented a novel algorithm to determine whether two route choices are identical. We find that for students using the same mode of transportation across all days, the percentage of subjects selecting the same route is very high, in the order of 94\%. We detail in the Methodology \ref{sec:meth-consistency} and Appendix \ref{sec:app-comproute} the algorithm used to obtain this number.

Finally, we identify a restricted set of clusters that have the property of being consistent throughout at least two days of experiment, i.e. the members of the cluster are the same in distinct days of the same week. Members may drop out of their cluster if their starting time or starting point are different from one morning to the next, or if they use another mode of transportation. We find that for these consistent clusters, close to 50\% of them have the property that the fastest individual on one day remains the fastest for all days where this cluster appears, showing again a certain degree of consistency in the population.

\subsection{Individual Optimality and Empirical Imitation-Regret}
\label{sec:regret}

To answer the question of individual optimality, we compare the durations of the morning trip for the subjects. A fair comparison is only achieved when looking at students leaving from the same neighborhood on the same day and at roughly the same time, going to the same school and using the same mode of transportation. The notion of neighborhood is expanded upon in our following methodology section, where we describe how the clustering of the data was achieved.

In the cases where the class of comparable subjects has more than two individual students, we collect the \textbf{empirical imitation-regret} encountered by every student in the class. To do so, we find the student in the class with minimal trip duration and set her imitation-regret to zero. For other members of the class, the empirical imitation-regret is equal to the (non-negative) difference between their trip duration and the minimal trip duration.

Our notion of empirical imitation-regret shares its name with the traditional regret measure, commonly found in the learning and multi-agent systems literature, for the following reason. The players here are faced with multiple strategies that they can choose from: the routes that go from their neighborhood to the destination. They may not know about current traffic conditions or which route will take the least amount of time but nevertheless have to make a decision. A posteriori, this decision can be compared with the best action they could have implemented on that day, and the difference is the imitation-regret. The appearance of the word ``imitation'' is due to the fact that we compare the decision solely with other players' choices of routes: a better route that is not used by any of the students in the cluster will therefore not be considered here. This drawback is shared with many natural learning dynamics and thus can be interpreted as a reasonable assumption on students' decisions.

\begin{figure}[!ht]
	\centering
	\begin{minipage}[c]{0.46\linewidth}
        \centering
        \includegraphics[width=0.95\textwidth]{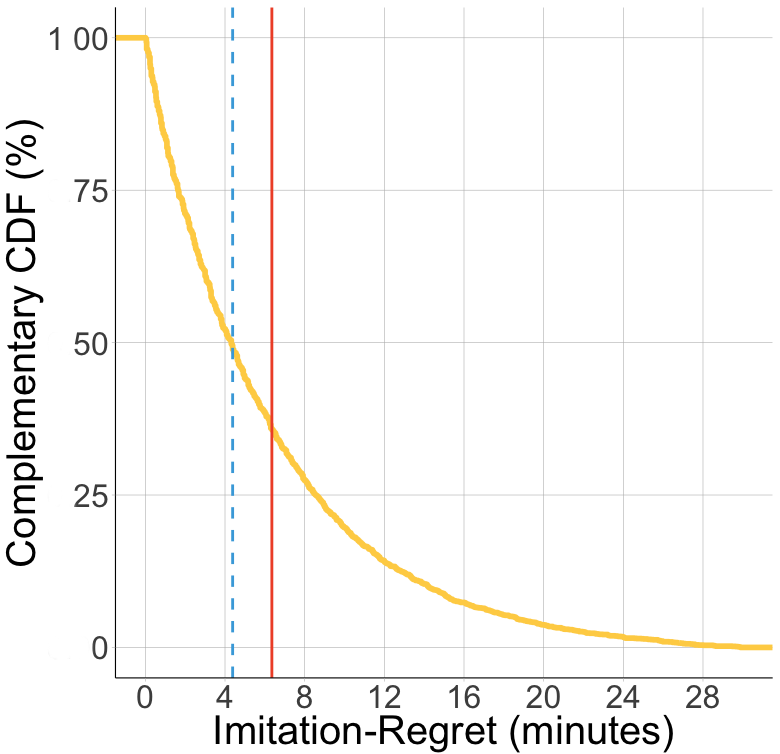}
	\end{minipage}
	\hspace{0.05\linewidth}
	\begin{minipage}[c]{0.46\linewidth}
		\centering
		\includegraphics[width=0.95\textwidth]{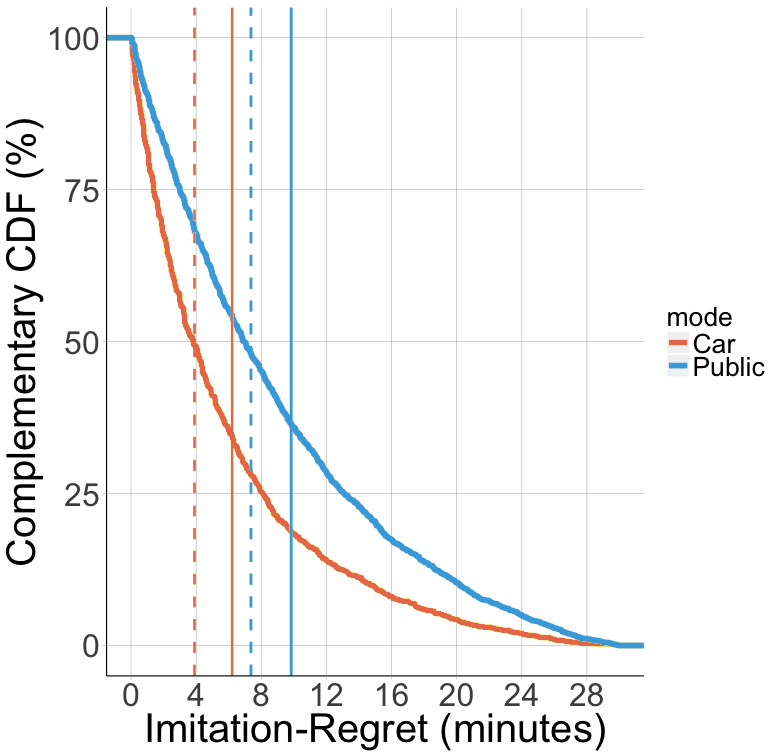}
	\end{minipage}
    \caption{\textit{Left:} Complementary cumulative distribution function of the imitation-regret. We aggregate all days of the experiment in a single figure and remove students with zero imitation-regret (in other words, the baseline students). The mean imitation-regret signalled by the red line is equal to 6 minutes, while the median imitation-regret plotted with the dashed blue line is equal to 4 minutes and 40 seconds. Sensitivity analysis results are presented at the end of this Section for different parameter values. \textit{Right:} Comparison of complementary CDF of imitation-regret per mode of transportation.}
    \label{fig:regretcdf}
\end{figure}

The measure of empirical imitation-regret depends naturally on the geographical area covered by the neighborhood. As the area increases, so does the accumulated imitation-regret, since the minimum is taken over a larger set of students. However, neighborhoods that are too large lose in precision, as two different subjects in the same cluster may have very different trip lengths. The results in this section use a geographical cluster size of about 400 meters, while we perform sensitivity analysis in the methodology to show the robustness of our findings.

Low empirical imitation-regret is a necessary condition for equilibrium. Indeed, at equilibrium, all comparable subjects should perform their trip in roughly the same amount of time. If one individual encounters a imitation-regret of say, 10 minutes, she may be better off by switching to a different route, e.g. the one used by the fastest individual in the cluster.

On the other hand, a high empirical imitation-regret warns us that some users are unable to find the fastest route to reach their destination. We see two possible directions to explore after such a conclusion. If we assume that individuals are solely interested in minimizing their trip duration --- perhaps a fair assumption for the morning trip, constrained by the hard deadline of the class start ---, then the network may benefit from the injection of information on how to traverse it. Otherwise, a high empirical imitation-regret reveals that other factors enter into consideration when the student is selecting the route, such as finding the least expensive one, the more climatised one or one that is shared with other students. The additional data collected by the sensor (e.g. temperature, proximity to other sensors) can indeed be articulated to uncover the nature of these factors.

In Figure \ref{fig:regretcdf}, we plot the complementary cumulative distribution of the emprirical imitation-regret. A point on the curve indicates which fraction of individuals (read on the \(y\)-axis) have empirical imitation-regret greater or equal than \(x\) (read on the \(x\)-axis). We also give the mean (solid red line) and median (dashed blue line) experienced empirical imitation-regret. It should be noted that the empirical imitation-regret distribution and its moments do not include the students for which the imitation-regret is zero, i.e. the best in the cluster.

Larger geographical cluster sizes give rise to larger average empirical imitation-regrets, but the results are relatively robust. The mean empirical imitation-regret oscillates around 6 minutes, while the median one is situated around 4 minutes and 40 seconds. This result motivates the introduction of a solution parametrised by two values, \( \epsilon \) and \( \delta \). The reported measurements constitute an \( (\epsilon, \delta) \)-equilibrium if we find that a fraction \( 1-\delta \) of users experience at most a quantity \( \epsilon \) of imitation-regret. The experiment yields values \( \epsilon = 22 \) minutes and \( \delta = 0.05 \).

Finally, we study the imitation-regret between modes, i.e taking the regret with respect to the fastest individual in the cluster, irrelevant of transportation mode. We focus our analysis on mixed clusters, where at least one individual using public transportation and one individual using private transportation appear. We have over 1,400 such clusters, and in close to 80\% of them, the fastest individual is a private transportation user. Over these 1,400 clusters, the average imitation-regret incurred by public transport users compared with the fastest private transportation user in their cluster is close to 8 minutes and 30 seconds. For the same population of bus and train users, the average duration of a trip is close to 25 minutes, indicating that the fastest car user spends roughly two thirds of this time to reach destination. Figure \ref{fig:regretcdf} plots the distributions of imitation-regret for the two classes of users.

\subsection*{Sensitivity Analysis of Clustering Methods}
\label{sec:app-sensanal}

\begin{figure}
	\centering
	\begin{minipage}[c]{0.46\linewidth}
        \centering
        \includegraphics[width=0.95\textwidth]{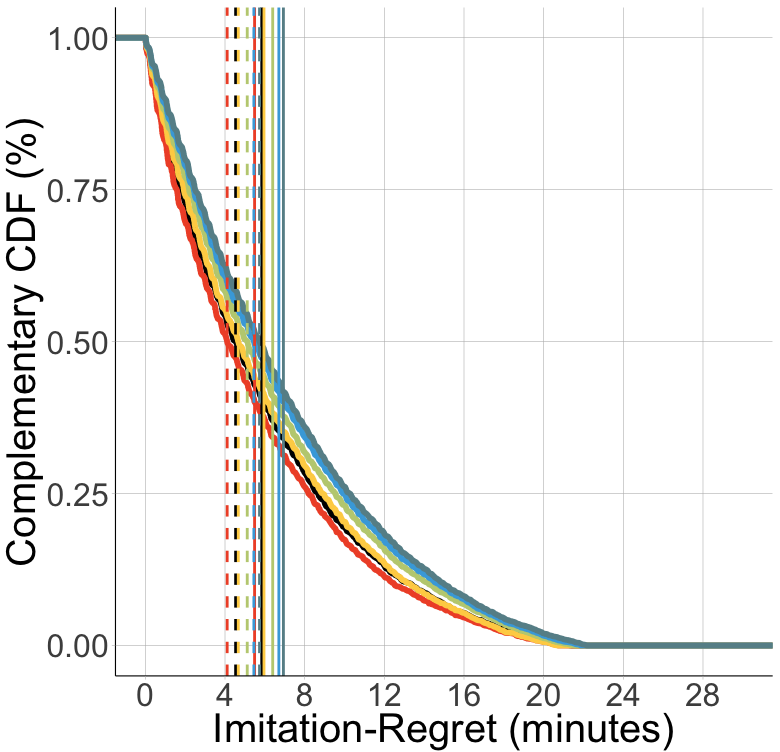}
	\end{minipage}
	\hspace{0.05\linewidth}
	\begin{minipage}[c]{0.46\linewidth}
		\centering
		\includegraphics[width=0.95\linewidth]{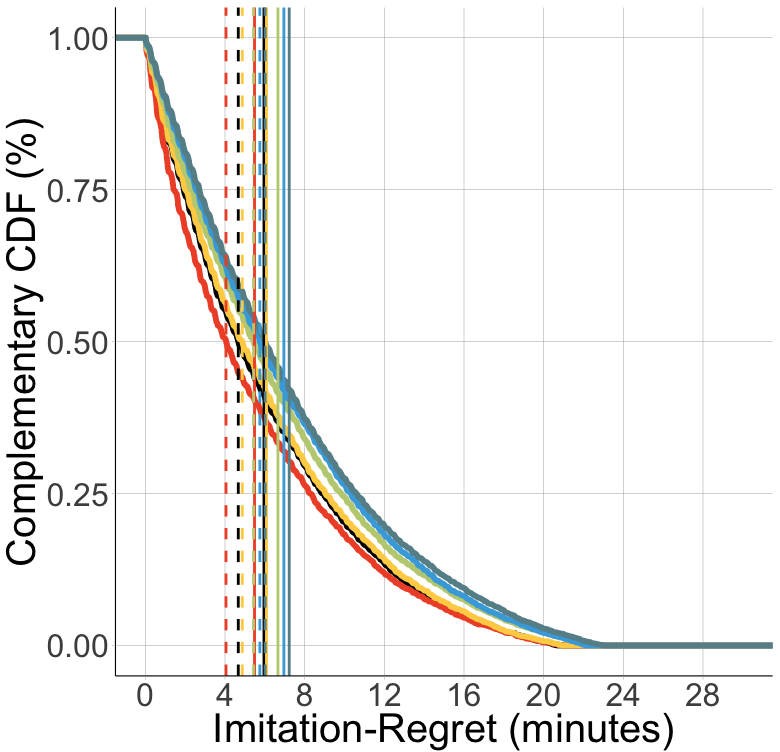}
	\end{minipage}
    \caption{The plots are similar to the one presented in the Findings section of the paper, with the difference that we are plotting the curves obtained from different clustering methods. \textit{Left:} We use as baseline the curve in black representing the results for the grid clustering with cell size 400 meters and plot in colors the curves obtained with different ball sizes (200, 400, 600, 800, 1000 meters, respectively red, yellow, green, blue and dark blue). Note that the means and medians increase as the size of the cluster grows. \textit{Right:} The baseline (in black) is now the curve for the ball clustering with 400 meters diameter and the colored curves correspond to the grid clustering with sizes (200, 400, 600, 800, 1000 meters, same colors).}
    \label{fig:regretcdfsens}
\end{figure}

We plot in Figure \ref{fig:regretcdfsens} the empirical complementary cumulative distribution function of regret in minutes. To obtain the curve, we aggregate all measures of imitation-regret and remove the baseline trips of zero imitation-regret from the data. The complementary cumulative distribution shows for a point \( (x, \overline{F}(x)) \) the fraction of trips \( \overline{F}(x) \) that have imitation-regret greater or equal to \( x \), in minutes.

\subsection{Societal Optimality and the Stress of Catastrophe}
\label{sec:poa}

The Stress of Catastrophe is introduced to give a measure of the weight of externalities in the system. As more agents join the road network, congestion increases on the links. Classically, the Price of Anarchy has been employed to quantify how bad the selfish decision-making of these agents affected the efficiency of the system, compared to the social optimum that a central planner can implement.

But estimating the social optimum of a system from the data is a perilous task. First, exact demands need to be known for every origin-destination pair of the agents. Second, latency functions for every edge of the network need to be estimated. Third, the global optimum flow maximizing the social optimum function needs to be computed. Additionally, the PoA does not fully capture the effects of a tragedy of the commons that congestion presents. In such a state, it is not costly for one additional individual to enter the system, but since all agents do so, the global welfare is very much diminished. But congestion can reach levels after which even the action of a central planner has little effect, yielding a low PoA that may not reflect the inefficiency at hand.

The Stress of Catastrophe eschews these pitfalls by providing an optimistic lower bound to the socially optimal trip durations. It stems from the simple fact that a crude lower bound to the optimal trip duration is one in which no one else is present on the road. Using Google Directions API, free-flow trip durations are obtained and give us a ``blue sky'' ideal lower bound. Comparing the actual recorded trip duration length to this lower bound in turn yields a ratio of how much faster the trip could have been in a no-externality scenario.

\begin{figure}
	\centering
	\begin{minipage}[c]{0.46\linewidth}
		\centering
	    \includegraphics[width=0.95\textwidth]{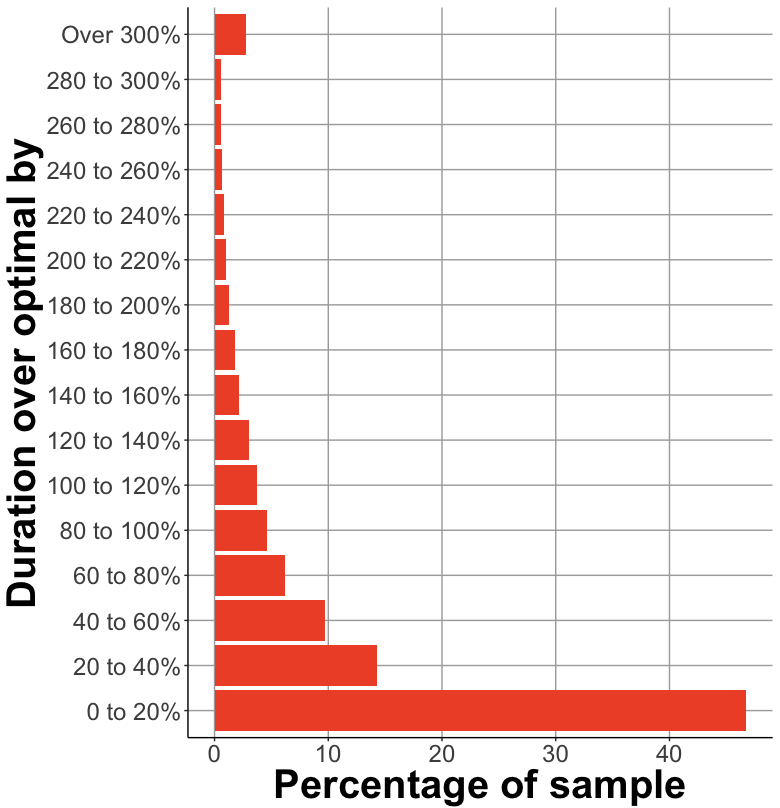}
	\end{minipage}
	\hspace{0.05\linewidth}
	\begin{minipage}[c]{0.46\linewidth}
		\centering
		\includegraphics[width=0.95\linewidth]{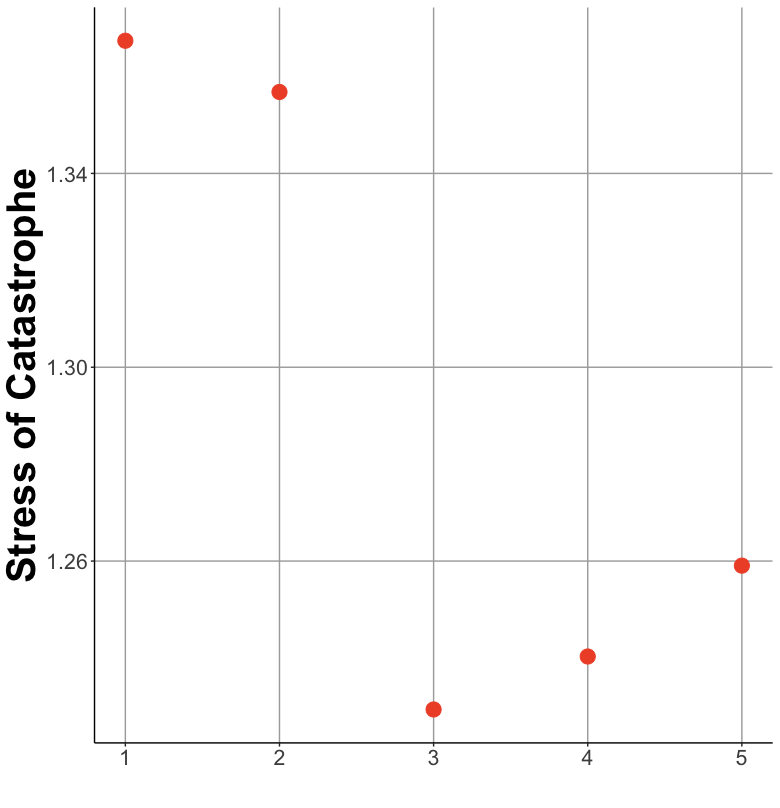}
	\end{minipage}
    \caption{\textit{Left:} Histogram of deviations from the free-flow optimal trip durations. \textit{Right:} Stress of Catastrophe computed across the five days with the highest record of unique students (sample size \(>\) 1,500). The values are between 1.23 and 1.37.}
    \label{fig:soc}
\end{figure}

Formally, we define the Stress of Catastrophe (SoC) from our data as such:
\[
    \texttt{\textbf{SoC}} = \frac{\texttt{\textbf{Cost}(Recorded trip duration)}}{\texttt{\textbf{Cost}(Trip durations (free-flow / light traffic))}}
\]
To give an idea of the measure in our dataset, we plot in Figure \ref{fig:soc} the histogram of percentages of deviations from the free-flow optimal trip duration. We see that most students are relatively close to this minimum bound while as the gap grows, fewer students are found.

Since the denominator is a lower bound to the socially optimal cost, we also have the following corollary:
\[
     \texttt{\textbf{PoA}} = \frac{\texttt{\textbf{Cost}(Recorded trip durations)}}{\texttt{\textbf{Cost}(Optimal trip durations)}} \leq \texttt{\textbf{Soc}}
\]

The question is now how optimistic is this upper bound? Our results show that SoC~\( = 1.34 \), when the SoC is computed with both car and transit users. But discriminating between the two yields a much more contrasted picture: the SoC for transit users is found to be 1.18, indicating that students using public transportation have little room to improve their trip duration. Conversely, the SoC varies significantly depending on the traffic conditions for subjects taking private transportation to school. In free-flow conditions, we find the SoC to be equal to 1.86. Details of the SoC for individual days of the experiment can be found in Figure \ref{fig:soc}.

It is remarkable that such an optimistic upper bound is however so close to 1. How does the PoA overestimate the inefficiency of the network then? Consider PoA results found in the literature, such as the \( 2.151 \) ratio of derived by \cite{Roughgarden09} in the case of degree 4 polynomial cost functions. The latest class is often used by network engineers to model the congestion on real roads, following the Bureau of Public Roads standard.

But the average estimated free flow time travel of the sample is 21 minutes. Assuming the SoC to be as large as the \( 2.151 \) bound, on average a commuter would spend~\( 2.151 - 1.34 = 0.811 \) times more in transit, i.e. 17 minutes more per commuter. In other words, pessimistic predictions of the PoA would entail a loss of over 730,000 hours per day, if we assume all of the 2,200,000 active individuals and 400,000 students were commuting on that day, a large mismatch with the actual system performance.

\section{Methodology}

To analyze the dataset, our algorithms make use of previously detailed procedures to extract semantic information from the raw stream of locations. We add further background and references for these algorithms in Appendix \ref{sec:app-semantic}.

\subsection{Consistency Between Trips}
\label{sec:meth-consistency}

Each student carries the sensor for up to 4 days in a week, allowing us to compare the morning trips taken by the same student between different days. Since the presence of noise in the sensor data and trip detection algorithm does not guarantee us that the whole week of experiment will be available, we filter out students for which only one morning trip is available. In our clean dataset, we have 15,875 individual students, out of which 9,352 have two or more trips logged in. On average, we have 2.44 trips per individual student. For these subjects, three analyses are carried out.

We first compare the modes of transportation selected by the student in the morning trips. Table \ref{tab:modeconsis} differentiates students that chose the same mode of transportation across all mornings from those that used multiple ones.

\begin{table}[!h]
\setlength\tabcolsep{6pt}
\centering
    \begin{tabular}{lcccc}
        \toprule
        & Metro & Bus & Car & \textbf{Total} \\
        \toprule
        Number of students & 1,278 & 1,448 & 3,070 & 5,796 \\
        \toprule
        Number of students (bus and metro grouped) & \multicolumn{2}{c}{3,060} & 3,070 & 6,130 \\
        \toprule
    \end{tabular}
    \caption{Number of students using consistently the same mode of transportation}
    \label{tab:modeconsis}
\end{table}

In a second and more granular analysis, we compare the routing decisions of the student at road level. Our aim is to determine whether the student selects the same route consistently to reach school in the morning. To achieve this task, we need a distance \( d_{a, b} \) measuring the similarity between two sequences \( a = (a_i)_{i=1}^{n} \) and \( b = (b_j)_{j=1}^{m} \) of coordinates. If the sequences are close, we can conclude that the same route has been selected on two occasions. We refer the interested reader to Appendix \ref{sec:app-comproute} for more information on how this distance is computed and tested.

\smallskip

Our third and last analysis is carried out on a subset of the data composed of 8,526 rows, one for each trip recorded that appears in a cluster containing of at least two trips. A filtered dataset of 7,674 rows is returned by removing trips with duration falling below or above the 5th and 95th percentile, respectively. This method allows for eliminating trips recorded with sensor error (e.g. not stopping the trip when the student reaches the school).

This filtered dataset contains 3,113 distinct clusters, i.e. unique sets of members. Of these, only 230 have the consistency property of featuring the same set of members for at least two days of experiment in the same week. This set is further reduced by taking only clusters where the individual with the shortest trip duration is also the fastest in every other day of the same cluster. As an example, if A and B belong to the same cluster on day 1 and day 2, and A's trip duration is shorter than B's on both day 1 and day 2, then the cluster is counted in our final dataset. We find that 113 such clusters are obtained at the end of the procedure, about half of the dataset of consistent clusters.

\subsection{Clustering and Empirical Imitation-Regret}
\label{sec:meth-imitregret}

We focus on the duration of the students' morning trip from home to school. To quantify how well the Singapore routing system performs, we obtain a lower bound of the total cost incurred by the students from the comparison between similar trips. More precisely, we divide the subjects in clusters and find for each cluster the student that reaches school in minimal time.

The clusters are indexed by 4 different variables:
\begin{itemize}
    \item \textbf{Geographical location \( l \):} Students living in the same neighborhood are grouped together. Results obtained for different cluster sizes $r$ are shown in Section \ref{sec:app-sensanal}.
    \item \textbf{Time of departure \( t \):} It is not accurate to compare a student departing from home at 6 am with one starting at 8 am. For this reason, students leaving on the same day within the same time frame are grouped together, using a window size of 20 minutes.
    \item \textbf{Destination \( s \):} Students going to the same school are grouped together. In the case that two or more schools share the same location (e.g. a Primary and a Secondary school), students attending either one of them are added to the same cluster.
    \item \textbf{Mode of transportation \( m \):} Two modes of transportation are discriminated between: private transportation (car, taxi) vs. public transportation (bus, train).
\end{itemize}

To obtain the geographical locations $l$, two spatial clustering methods are implemented. In the first version, we find the smallest bounding box that contains all the home locations of the students. We divide this bounding box in cells of equal edge size \( r \), e.g. \( r = 400 \) meters, and assign to the same geographical clusters students with home locations inside of the same cell. This is a grid-based method that partitions the space into a finite number of cells from a grid structure. Its main advantage is its fast processing time. Sensitivity analysis measures are reported for different cell sizes.

The second version of the spatial clustering approach is based on a distance rule where all home locations of the students in the same cluster should be within $r$ meters of each other. This is a hierarchical clustering method using decision trees based in the geodesic distance matrix of all trips. This technique, although computationally more expensive, ensures that the distance rule holds for all the trips.

Figure \ref{fig:clusters} (also Figure \ref{fig:clusters-big} in Appendix for a larger version) shows the visual comparison of the two different spatial clustering methods for $r=400$ meters for all 15,875 students considered in our study. The red dots represent the home locations and each dot corresponds to exactly one student. The blue triangles show the school locations of the 89 schools\footnote{It is interesting to note that some students have home location in Malaysia, and commute from Malaysia to Singapore daily for study.}. The grid-based method (top left) is a simple but efficient strategy, simply counting the points that fall in specific cells of the mesh. On the other hand, the distance rule approach (top right) can be visualized by circles of diameter equal to 400 meters.
Inside each circle, the maximum distance between any two home locations is 400 meters. The algorithm optimizes a criterion function and the centroid of each spatial cluster can be easily identified, making it a powerful method to build the clusters. Recall that Figure \ref{fig:clusters} is presented only for visualization purposes since inside each spatial cluster (cells/circles), students might be mixed among different transportation modes and different destination schools.

The value of 400m is picked for the following reasons. First, assuming a uniformly random distribution in the cell their expected distance would be a little over 200. In practice this distribution is concentrated on blocks of flats and two students could easily be living in the same block. Let's assume that the students are at distance of 200m. There are two cases, either both the students drive/are driven or they take bus/metro. If they drive, this distance is noise. If they use metro/bus then since they go to the same school, they typically would use the same bus/metro and board it at the same regular time. The only differentiating factors are the difference in the distance they cover on foot (in a geometric world via triangle inequality less than 200m) and the difference between the amounts of buffer time (arrive a little earlier) at the stop. The rest of the route is identical.

We obtain a set of clusters \( \{ C_{l, t, s, m} \}_{l, t, s, m} \) where each \( C_{l, t, s, m} \) contains the trip durations \( t_{l, t, s, m}^i \) of students in the cluster. If several students belong to the same cluster, we find the student whose trip has the minimum duration among all trips in the cluster. We call this trip the \textit{baseline} \( t^b_{l, t, s, m} \), against which the remaining trips will be compared.

We next define the imitation-regret for student \( i \) in cluster \( C_{l, t, s, m} \) by
\[
    R^i_{l, t, s, m} = t^i_{l, t, s, m} - t^b_{l, t, s, m}
\]
Obviously the imitation-regret for the baseline student is zero, and non negative for everyone else. We are interested in seeing how large the deviations from the baseline can be, as a necessary condition for the system to be at equilibrium is that these deviations must be close to zero. Sensitivity analysis is performed to account for the parameters used in the clustering method. We present it in Section \ref{sec:app-sensanal}.

\subsection{Estimating the Stress of Catastrophe}

This section offers more details on how the lower bound to the trip durations is obtained. The Google Directions API is queried for the best route and minimal trip duration, for both private vehicle transportation as well as public transit.

For car users, the API is called was called using an ``optimistic'' parameter, specifying that best case trip durations ought to be returned. Collection was made throughout several days in February 2017. Thereafter, the minimum between all returned durations and the student's actual trip duration was used as a ``free-flow'' estimate.

For public transportation users, the API is not time-dependent, i.e. will not return results that depend on the congestion. To obtain a best cast trip duration, we remove potential waiting time at MRT or bus stations. Collection was made parallel to that of private transportation trips.

To minimize the number of requests to the API, we employ the grid clustering method described in Section \ref{sec:meth-imitregret} and query for the best route between the cluster centroid and a school, if some of its students' homes are located inside the cluster. Some of the requests do not return satisfactory results, either due to a Not Found error or when repositioning the starting point of the trip too far from the student's home: these are dropped in the analysis. We refer to Table \ref{tab:optimal} for the number of datapoints collected for the classes of users taking either public or private transportation.

\begin{table}
    \setlength\tabcolsep{6pt}
\centering
    \begin{tabular}{lcc}
        \toprule
        & Number of points & Mean optimal trip duration (mins) \\
        \toprule
        Private transportation & 8,745 & 11'49 \\
        \toprule
        Public transportation & 11,697 & 28'18 \\
        \toprule
        Total & 20,442 & 21'04 \\
        \toprule
    \end{tabular}
    \caption{A few statistics on the collected optimal trips}
    \label{tab:optimal}
\end{table}

\section{Connections to Other Work}
\textit{Algorithmic Game Theory and Econometrics.} Recently there has been a surge of interest in combining techniques from algorithmic game theory with the traditional goals of econometrics \citep{bajari2013game, syrgkanis2015algorithmic}. These works  employ a data-driven approach to analyzing the economic behavior of real world systems and agent interactions.
In \citep{Nekipelov:2015:ELA:2764468.2764522} the authors developed theoretical tools for inferring agent valuations from observed data in the generalized second price auction without relying on the Nash equilibrium assumption, using behavioral models from online learning theory such as regret-minimization. They apply their techniques on auction data to test their effectiveness.

Following this work, \citep{DBLP:journals/corr/JalalyNT17} studies the behavior of real housing market agents based on data from an online bidding platform. The results inform the design of the auction platform and point towards data-driven policies helping the agents make decisions. The latter idea is made more explicit in a recent article by some of the authors \citep{Nekipelov2017:IAD:3116227.3035966}. In a sense, our present work also advocates using data to gauge users interactions but our focus is on routing games, for which it is harder to gather sanitized data. Furthermore, we develop new metrics that are more informative about the state of the system than the price of anarchy.

In \citep{hoy2015robust} the authors provided tools for estimating an empirical PoA of auctions. The PoA is defined as the worst case efficiency loss of any auction that could have produced the data, relative to the optimal. However, auctions and routing games each pose a totally distinct set of challenges. In our setting, the problem of translating data streams to game theoretic concepts adds a rather nontrivial layer of complexity. For example, even identifying the action chosen by each agent, i.e. their routes, is tricky as it requires to robustly map a noisy stream of transportation data into a discrete object, a path in a graph. It should be noted however that our notion of Stress of Catastrophe provides an upper bound on the empirical PoA, as detailed in Section \ref{sec:poa}.

\textit{Price of Anarchy for Real World Networks.} One earlier paper tangentially connected to estimating the PoA of congestion games is
\citep{buriol2011smoothed}. This is a theoretical paper that provides PoA bounds for perturbed versions of congestion games.
As a test of their techniques, they heuristically approximate the PoA  on  a few benchmark instances of traffic networks available for academic research from the Transportation Network Test Problems~\citep{BarGera} by running the Frank-Wolfe algorithm on them. No experiments were performed and no measurements were made. Naturally, this approach cannot be used to test PoA predictions, since it presumes that PoA reflects the worst case possible performance and then merely tests where do these constants lie for non-worst case routing networks.

In effectively parallel independent work \citep{zhang2016price, zhang2016data} focused on quantifying the inefficiencies incurred due to selfish behavior for a sub-transportation network in Eastern Massachusetts, US.
They use a dataset containing time average speed on road segments and link capacity in their transportation sub-network.
The authors estimate daily user cost functions as well as origin-destination demand by means of inverse optimization techniques using this dataset.
From this formulation they compute estimates of the PoA, whose average value is shown to be around $1.5$.
In contrast to their approach our dataset contains detailed individual user information, which allows for estimates not only of systemic performance but also of individual optimality (e.g. imitative-regret) as well as test to what extent is the system indeed near stasis (i.e. in equilibrium). Also, their approach does not capture how bad the resulting traffic is, i.e., the tension between Price of Anarchy and Tragedy of the Commons, whereas our approach addresses both. Finally, our estimations are derived from explicit online measurements of the system performance and are not reverse engineered by estimating user cost functions which inevitably introduce new errors that cascade through all the calculations.

\textit{Transportation Science and Game Theory.} The Braess paradox famously asserts that adding a road to a congested traffic network could increase the overall journey time, and in effect, increase the PoA of the network. It is also known to be indeed a prevalent phenomenon in real traffic networks~\citep{steinberg1983prevalence}. In \citep{bagloee2014heuristic}, the authors examine heuristics to identify roads in traffic networks whose removal would  improve the overall system performance, applying the Braess paradox in the reverse direction. This is an interesting example where a game theoretic phenomenon inspires a possible improvement to a real-world transportation problem.


%


\section{Discussion}

This is hopefully not the end but the beginning of a thorough experimental investigation into the rich game theoretic literature of  routing games. Clearly, there are many open questions and challenges to be addressed. Although it does not seem realistic to expect to have a global concurrent view of the movement of all commuters in any real life traffic network (especially one of the size of Singapore), it would be helpful to develop more computational tools and techniques that allow us to increase further the sample size. Naturally, it would be rather useful to compare and contrast the state of traffic networks in different parts of the globe from a game theoretic lens. Specifically, to what extent is Singapore's routing efficiency an artifact of local policies, cultural norms and traffic topology and which of these estimates are relatively robust across many cities and cultures? Indeed, there is plenty of work to be done from the standpoint of behavioral game theory to understand the validity of common game theoretic assumptions/models (e.g. equilibration).

 In the opposite direction we also hope that our work might encourage new theoretical investigations into system efficiency as well. We believe that a particularly interesting direction has to do with understanding the interconnection between Price of Anarchy effects and Tragedy of the Commons behaviors. It would be practically important to understand which game theoretic settings are particularly sensitive to a over-demand type of catastrophe where more users keep entering the system lured in by the effects of a low and decreasing Price of Anarchy. We hope that our notion of stress of catastrophe might work as a helpful starting point in this direction. 





\bibliographystyle{apalike}
\bibliography{references,sigproc2,sample-bibliography,sigproc_ec13}

\newpage

\section*{Appendix}
\label{sec:appendix}
\subsection{The National Science Experiment}
\label{sec:app-nse}

As part of the Smart Nation programme, the \emph{National Science Experiment} (NSE) has the primary goal of inspiring future generations of students to pursue technical education. The NSE is a nationwide project in which over 90,000 students from primary, secondary and junior college wore a sensor, called SENSg, for one week in 2015 and 2016. The SENSg sensors collect ambient temperature, relative humidity, atmospheric pressure, light intensity, sound pressure level, and 9-degree of freedom motion data. In \citep{Wilhelm2016}, a detailed technical description of the sensor is presented. The design of the SENSg was created with the specific requirements to have a low-cost device for a one-week crowd-sensing experiment, that does not need to be charged. It led up to the mass-production of 50,000 sensor nodes.

The SENSg scans the Wi-Fi hotspots which are used to localize the sensor nodes as well as to move sensor data to a back-end server. All environment values are sampled every 13 seconds using the Wi-Fi based localization system. The raw collected datapoints are then post-processed to obtain semantic data, employing state-of-the-art methods described in \citep{Monnot2016, Wilhelm2017, Zhou2017}. The semantic data covers the identification of individual trips within the discrete stream of locations, inference of the activity performed at each endpoint and transportation mode classification. We give a brief summary of the first and third algorithms in Appendix \ref{sec:app-semantic}.

The NSE 2016 dataset contains data of 49,526 students who wore the SENSg sensor. This work uses the trip identification algorithm developed in \citep{Zhou2017} where five different modes can be identified, namely: (a) stationary; (b) walking; (c) riding a metro; (d) riding a bus; and (e) riding a car. With additional information from the LTA, the algorithm detection covers 8 metro lines, 106 metro stations, 260 bus services and 4,684 bus stops. Similarly, the 164 km of expressways and the 698 km of arterial roads in Singapore feed the algorithm to distinguish whether a subject is traveling in a car.

The experiment was designed to analyze homogeneous users, i.e. primary, secondary or junior college students, reucing the complexity of understanding mobility patterns. This work focuses primarily on morning travels of students who get to their schools from their homes. Even though the walking mode can be detected by the algorithm, we restrict the study to metro, bus, and car trips. To ensure the quality of our empirical results, we perform a strict data cleaning process, and the top and bottom 5 percents of the trip duration/distance traveled values were dropped to remove outlier points.

Table \ref{tab:dataset} charts the basic description of the dataset used in this study. A total of 32,588 clean trips are considered, covering 15,875 students and 89 schools. The number of students by school type is approximately equally distributed, hence capturing the routing behavior of students over a large space in Singapore. A few pointed insights can be derived from the table. For example, around 55 percent of the Secondary school students take public transport (metro or bus) to reach their schools, and this proportion increases to 75 percent for junior college students. In the case of primary school students, a large fraction of the trips are made by car, a number similar to that of the few adult Singapore residents who use car to reach their workplaces (21.9 percent of the residents according to the LTA reports).

\begin{table}
\centering
\begin{tabular}{lrrrrrr}\toprule
                                & \multicolumn{3}{l}{Mode (\% of the trips)}                       & \multicolumn{3}{c}{Total}                                                              \\ \toprule
\multicolumn{1}{l}{School type} & \multicolumn{1}{c}{Metro} & \multicolumn{1}{c}{Bus} & \multicolumn{1}{c}{Car}   & \multicolumn{1}{c}{Trips} & \multicolumn{1}{c}{Students} & \multicolumn{1}{c}{Schools} \\ \toprule
Primary                         & 5.50                     & 35.20 & 59.30                     & 9,373                     & 4,553 & 28                           \\
Secondary                       & 14.13                     & 40.80                   & 45.07                     & 10,789                    & 5,229 & 20                          \\
Junior College                  & 53.60                     & 22.14                   & 24.26                     & 12,426                    & 6,093                        & 41                          \\ \toprule
                                & \multicolumn{1}{l}{}      & \multicolumn{1}{l}{}    & \multicolumn{1}{r}{\textbf{Total}} & 32,588 & 15,875                       & 89             \\ \toprule
\end{tabular}
\caption{Dataset description}
\label{tab:dataset}

\end{table}

\subsection{Overview of Singapore's Routing Network}
\label{sec:app-singapore}

\begin{figure}
    \centering
    \includegraphics[width=.8\textwidth]{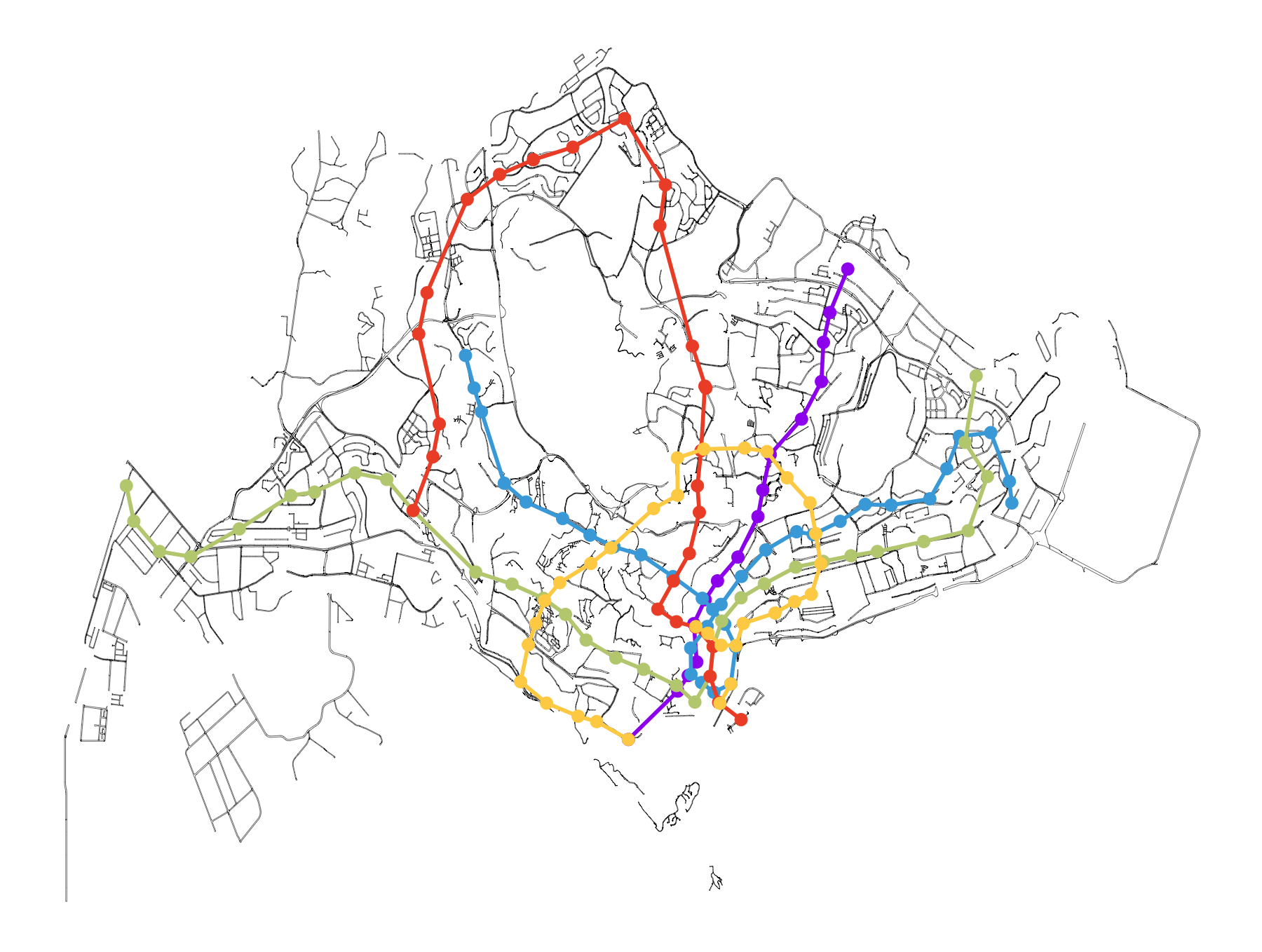}
    \caption{A plan of Singapore with major roads and MRT (Mass Rapid Transit) stations, by line.}
    \label{fig:singaporemap}
\end{figure}

Singapore is an island city-state located at the southern tip of Peninsular Malaysia in South East Asia (depicted in Figure \ref{fig:singaporemap}). According to the General Household Survey 2015 elaborated by the Department of Statistics Singapore, the city-state is home to 5.6 million people, with only an area of 712 km$^{2}$. Metro and bus are the two main modes of public transportation in this densely populated country. Private modes of transportation are mainly passenger cars like taxis or ride-hailing services and private vehicles.

The 2015 Annual Report published by the Land Transport Authority (LTA) indicates that the metro system consists of a network of 142 stations that can be grouped into 5 Mass Rapid Transit (MRT) lines and 3 Light Rail Transit (LRT) lines. This rail system moves more than 2 million passengers daily, and is the backbone of Singapore's public transport network and spans some 150km across the island. Metro operation hours are from 5:50 am to about midnight daily with a 2 to 3 minutes frequency during peak hours (7 am to 9 am). The studies in \citep{Sun2012, Poonawala2016} make an effort to understand crowds within the Singapore metro network. Singapore's public buses serve local transport within a town to the hub whereas the metro network is used for longer distance trips. The bus network is larger, with 260 bus services and 4,684 bus stops according to the 2015 Annual Report of the LTA. During peak periods, all buses are scheduled to arrive every 15 minutes or less with at least half of these scheduled to arrive every 10 minutes or less.

The metro and bus fares are paid using a contactless smart card, named EZ-Link, where users just tap the card to scan their train ticket at the gantries when commuters enter and exit the metro platform area or bus stop. The LTA regulates and oversees metro and bus transport, ensuring they meet safety standards. The rich information generated by the use of the EZ-Link card was exploited in the works of \citep{Sun2012, Lee2014,Holleczek2015,Poonawala2016}, attempting to find interesting insights for public transit planners.

On the other hand, the rapid development and the population growth of Singapore have led to an increase in the car usage. Still, in a small city-state like Singapore, the expansion of the road network is a constraint. The trade-off between the use of land for roads  (currently accounting for 12\% of the land) and other activities has become an important issue for planning authorities. With this in mind, the Government has put in place a combination of ownership and usage measures to manage road traffic. Private car users have to purchase a paper license that in many cases is more expensive than the car itself. Despite the economic strategy to regulate the car ownership, the 2012 Household Interview Travel Survey found that about 39\% of the trips in 2012 were made by private transport. Moreover, 46\% of households owned cars in 2012, compared to 40\% in 2008, and 38\% in 2004.

\subsection{From Raw Data to Semantic Data: Mode Identification}
\label{sec:app-semantic}

As previously mentioned, the SENSg sensor is equipped with a Wi-Fi module, which scans the information of surrounding Wi-Fi hotspots (MAC addresses, Service Set Identifier, etc.) every 13 seconds. Each 13-second interval of collected information is called a working cycle. Once the data is uploaded to the server, the Wi-Fi hotspots information is converted to estimated latitude and longitude. As explained in \citep{Zhou2017}, a first cleaning process is carried out to reduce the noise in the raw data. Then, each package of 100 working cycles is considered as a single sample, and the transportation mode identification algorithm starts by classifying the sample into:
\begin{itemize}
\item Vehicle samples: Samples that belong to vehicle trips, \textit{or}
\item Non-vehicle samples: Samples that can be either walking or stationary.
\end{itemize}

After smoothing the data by a heuristic approach, vehicle samples and non-vehicle samples are processed separately. We refer the interested reader to \citep{Zhou2017} for a detailed description of the treatment of non-vehicle samples. In the following, our intention is to provide a brief overview of how vehicle samples are processed. The properties of this particular algorithm are crucial to this work, since we endeavor to estimate accurate measures of regret by taking into account the best clean subset of pre-processed information available.

The key idea behind the algorithm is to conduct type classification on segments instead of individual data samples. There are two main parts in this algorithm: first, catch segments and second, classify them. In the first part (\emph{catching}), the vehicle segment detection works by using information about a set of parameters such as the velocity and acceleration of the sample. Note that at this stage some segments can belong to walking or stationary modes. In the second part (\emph{classifying}), all detected vehicle segments are classified into one of the transportation modes. By following 17 segment-wise features such as the distance between segment's start/end location to the nearest bus/metro stop, etc. this goal can be reached. Hence, the vehicle trips can be properly classified since it is possible in each sample to know how close public transportation lines are, and how long/fast the vehicle segment is. Car mode trips are segments that do not fall into a metro or bus classification, as returned by a Random Forest classifier.

In short, the algorithm first catches the potential vehicle mode trips and then classifies each of them into riding a metro, riding a bus or riding a car. A high level accuracy of 85\% is guaranteed, indeed superior to similar recent studies (e.g. \citep{Sankaran2014, Shin2015} or \citep{Zhu2016}).

\subsection{Comparing Routes}
\label{sec:app-comproute}

The distance \( d_{a, b} \) is obtained with an estimation of the area enclosed by the polygon formed by the concatenation of \( a \) and \( b \). More formally, since  \( a_1 = b_1 = \text{ Home} \) and \( a_n = b_m = \text{ School} \), we consider the area enclosed by the polygon \( (a_1, a_2, \dots, a_n, b_{m-1}, b_{m-2}, \dots, b_1) \). The distance gets more precise as the number of datapoints logged along the trip increases, although it is still possible to achieve good results for very sparse trips.

\begin{figure}[h!]
	\centering
	\begin{minipage}[c]{0.46\linewidth}
		\centering
		\includegraphics[width=0.95\linewidth]{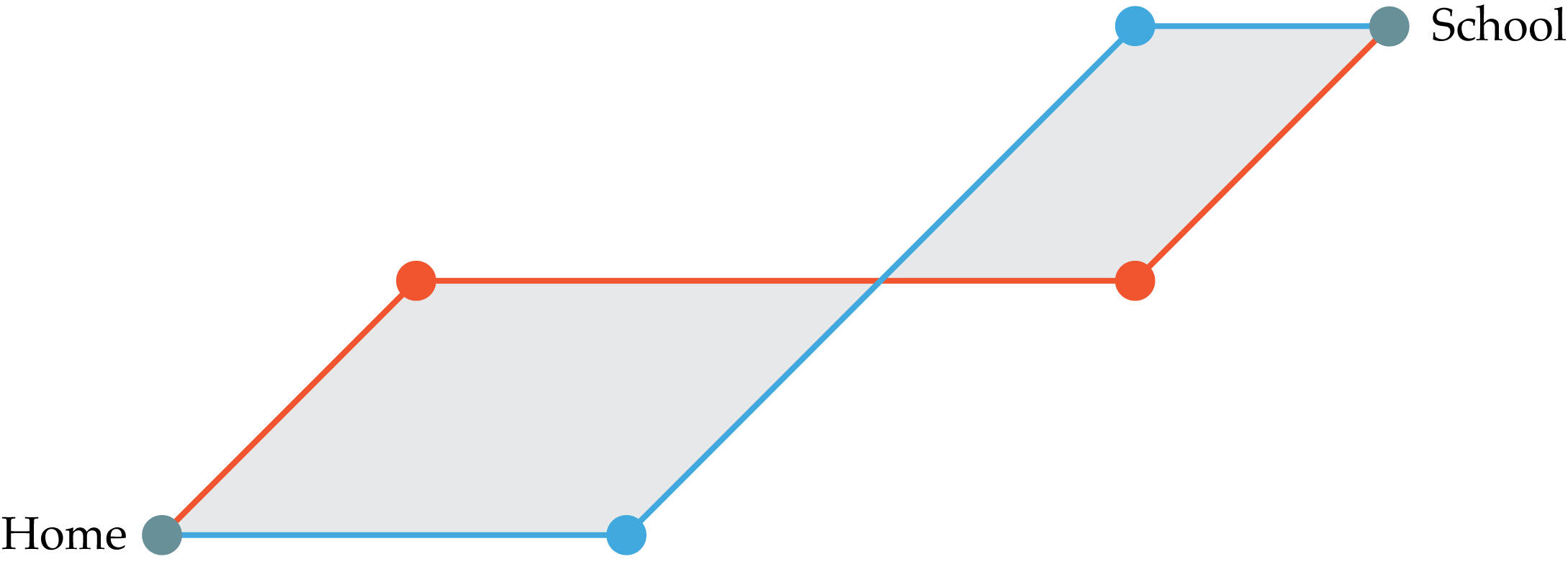}
		\caption{Two trips (red and blue) are plotted, with endpoints in dark blue. The distance between the trips is measured by the area of the darkened surface.}
		\label{fig:polygontrip}
	\end{minipage}
	\hspace{0.05\linewidth}
	\begin{minipage}[c]{0.46\linewidth}
		\centering
		\includegraphics[width=0.95\linewidth]{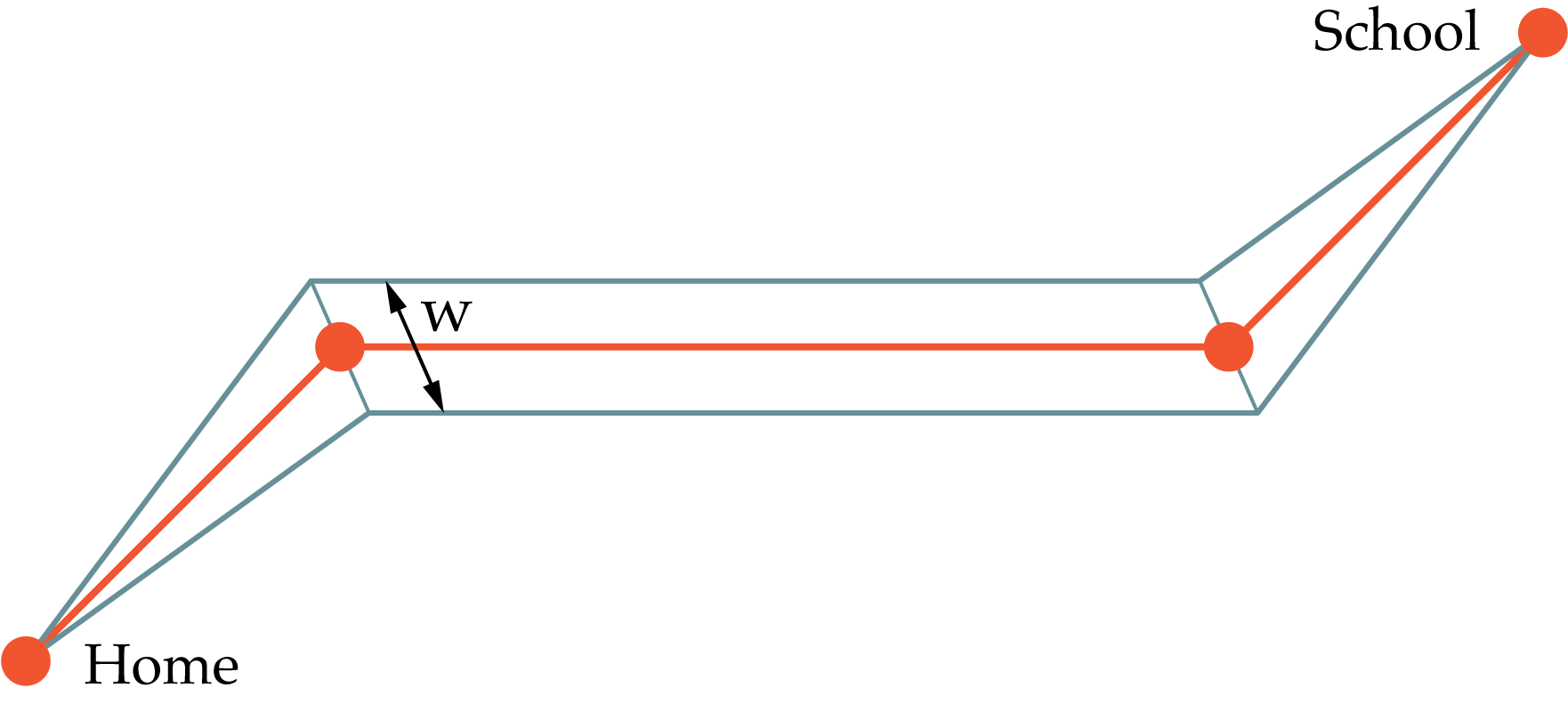}
		\caption{A four-point trip is plotted in red. The outer contour is obtained by fixing a band width \( w \).}
		\label{fig:outercontour}
	\end{minipage}
\end{figure}

We need a criterion to decide whether the previously obtained area is sufficiently small for the two sequences of coordinates to be considered consistent. To this end, we construct the outer contour of each sequence \( a \) and \( b \), defined by a polygon containing the sequence of coordinates. Intuitively, we construct a band around the stream of locations, and the area of that band allows us to determine what constitutes an acceptable deviation. We show in Figure \ref{fig:outercontour} a representation of the outer contour for a trip with 4 points. Given \( d^o_a \) and \( d^o_b \), respective areas of the outer contour for \( a \) and \( b \), we use the following criterion to classify the routes \( a \) and \( b \) as consistent:

\[
    d_{a, b} < \frac{d^o_a + d^o_b}{2} \Rightarrow a \text{ and } b \text{ are consistent.}
\]
The average is taken to ensure that both trips are considered equally in our criterion. Indeed, without the average, the algorithm may not be able to recognize a small deviation if only the outer contour of the shortest path appears on the right-hand side.

A parameter \( w \) controls the width of the band. If we set \( w \) to a value that is too large, we run the risk of incorrectly classifying different trips as consistent. On the other hand, a \( w \) that is too small may mark as non-consistent trips that make use of the same route. We have set the value of \( w \) by creating negative examples, comparing a trip with translated versions of the same sequence of coordinates. Visual analysis further confirmed the validity of its choice.

\begin{figure}
  \centering
  \includegraphics[width=0.8\linewidth]{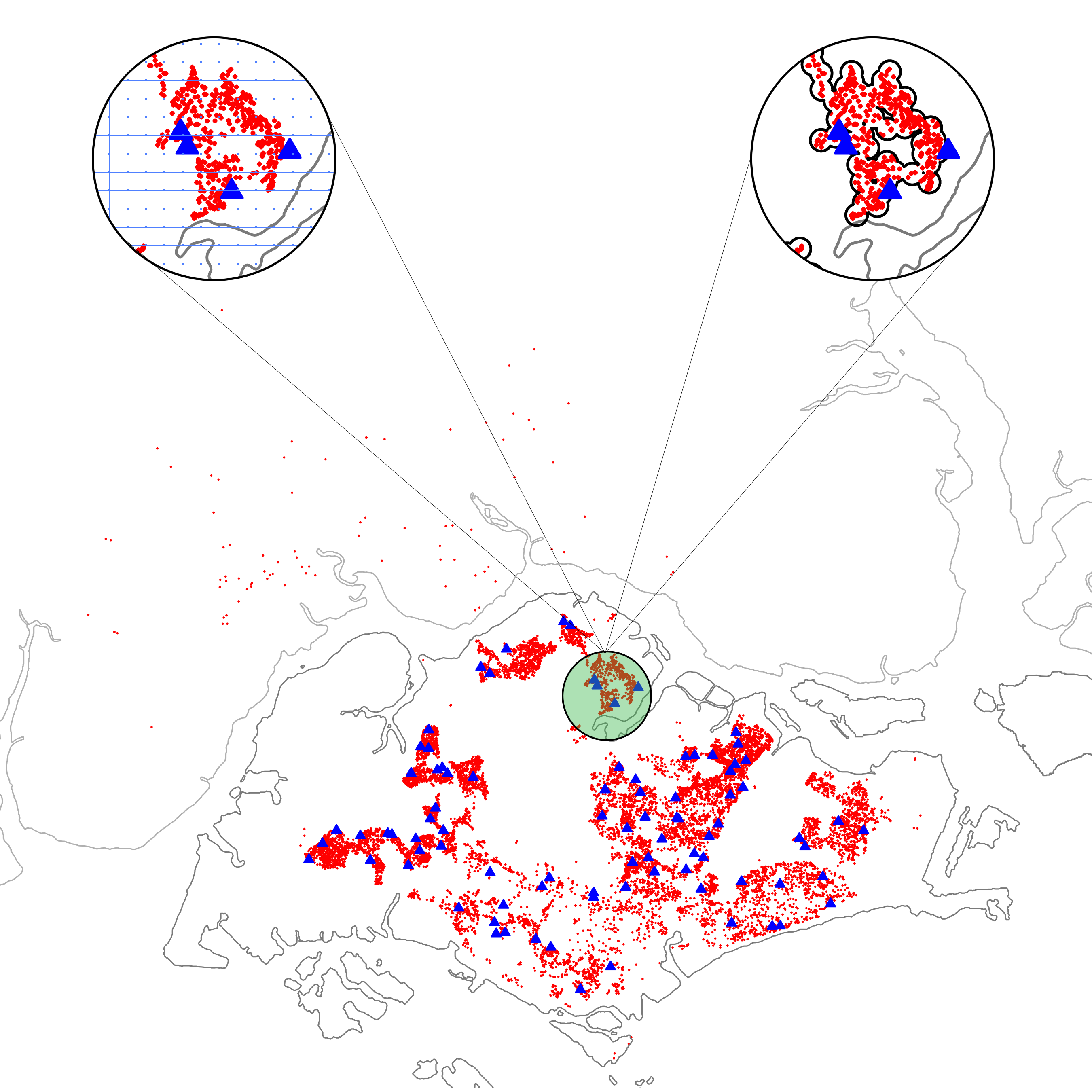}
  \caption{Home locations (red dots), school locations (blue triangles) and spatial clustering methods, discussed in Sections \ref{sec:regret} and \ref{sec:meth-imitregret}}
  \label{fig:clusters-big}
\end{figure}

\section*{Acknowledgements}
The authors would like to thank the National Science Experiment team at SUTD for their help: Garvit Bansal, Sarah Nadiawati, Hugh Tay Keng Liang, Nils Ole Tippenhauer, Bige Tunçer, Darshan Virupashka, Erik Wilhelm and Yuren Zhou. The National Science Experiment is supported by the Singapore National Research Foundation (NRF), Grant RGNRF1402.

Barnab\'e Monnot would like to acknowledge the SUTD Presidential Graduate Fellowship.
Georgios Piliouras would like to acknowledge
SUTD grant SRG ESD 2015 097 and MOE AcRF Tier 2 Grant  2016-T2-1-170.
Part of the work was completed  while Barnab\'e Monnot and Georgios Piliouras  were visiting scientists at the Simons Institute for the Theory of Computing (Economics and Computation Semester, Fall '15).

\end{document}